\documentclass[twocolumn]{aastex631}

\newcommand{\msun}{$M_{\odot}$}
\newcommand{\mstar}{$M_{\star}$}
\newcommand{\mdmhalf}{$M^{50}_{\rm DM}$}
\newcommand{\mhalo}{$M_{\rm halo}$}
\newcommand{\rhalo}{$r_{\rm halo}$}
\newcommand{\mvir}{$M_{\rm vir}$}
\newcommand{\rvir}{$r_{\rm vir}$}
\newcommand{\rhalf}{$r_{50}$}

\newcommand{\drhalf}{$\Delta \, \log \, r_{50}$}
\newcommand{\ddm}{$\Delta \, \log \, M^{50}_{\rm DM}$}
\newcommand{\dmhalo}{$\Delta \, \log \, M_{\rm halo}$}

\usepackage{hyperref}
\usepackage{amsmath}

\begin{document}

\title{Effects of galactic environment on size and dark matter content in low-mass galaxies}

\author[0000-0002-5908-737X]{Francisco J. Mercado}
\affiliation{Department of Physics \& Astronomy, Pomona College, Claremont, CA 91711, USA}
\affiliation{TAPIR, California Institute of Technology, Pasadena, CA 91125, USA}

\author[0000-0002-3430-3232]{Jorge Moreno}
\affiliation{Department of Physics \& Astronomy, Pomona College, Claremont, CA 91711, USA}
\affiliation{Carnegie Observatories, Pasadena, CA 91101, USA}

\author[0000-0002-1109-1919]{Robert Feldmann}
\affiliation{Institute for Computational Science, University of Zurich, Zurich CH-8057, Switzerland}

\author{Marckie Zeender}
\affiliation{Department of Physics \& Astronomy, Pomona College, Claremont, CA 91711, USA}

\author[0000-0003-1896-0424]{Jos\'e A. Benavides}
\affiliation{Department of Physics and Astronomy, University of California, Riverside, CA 92507, USA}

\author[0000-0003-1661-2338]{Joanna M. Piotrowska}
\affiliation{Cahill Center for Astronomy and Astrophysics, California Institute of Technology, Pasadena, CA 91125,USA}

\author[0000-0002-2762-4046]{Courtney Klein}
\affiliation{Department of Physics and Astronomy, University of California Irvine, Irvine, CA 92697, USA}

\author{Coral Wheeler}
\affiliation{Department of Physics and Astronomy, California Polytechnic State University Pomona, Pomona, CA 91768, USA}

\author[0000-0003-2806-1414]{Lina Necib}
\affiliation{Department of Physics and Kavli Institute for Astrophysics and Space Research, Massachusetts Institute of Technology, 77 Massachusetts Ave, Cambridge,
MA 02139, USA}

\author[0000-0003-4298-5082]{James S. Bullock}
\affiliation{Department of Physics and Astronomy, University of California Irvine, Irvine, CA 92697, USA}

\author[0000-0003-3729-1684]{Philip F. Hopkins}
\affiliation{TAPIR, California Institute of Technology, Pasadena, CA 91125, USA}
\affiliation{Cahill Center for Astronomy and Astrophysics, California Institute of Technology, Pasadena, CA 91125,USA}


\begin{abstract}

We utilize the cosmological volume simulation, FIREbox, to investigate how a galaxy’s environment influences its size and dark matter content. Our study focuses on approximately 1,200 galaxies (886 central and 332 satellite halos) in the low-mass regime, with stellar masses between $10^6$ to $10^9$ \msun. We analyze the size-mass relation (\rhalf\ - \mstar), inner dark matter mass-stellar mass (\mdmhalf\ - \mstar) relation, and the halo mass-stellar mass (\mhalo\ - \mstar) relation. At fixed stellar mass, we find that galaxies experiencing stronger tidal influences, indicated by higher Perturbation Indices (PI $>$ 1) are generally larger and have lower halo masses relative to their counterparts with lower Perturbation Indices (PI $<$ 1). Applying a Random Forest regression model, we show that both the environment (PI) and halo mass (\mhalo) are significant predictors of a galaxy's relative size and dark matter content. Notably, because \mhalo\ is also strongly affected by the environment, our findings indicate that environmental conditions not only influence galactic sizes and relative inner dark matter content directly, but also indirectly through their impact on halo mass. Our results highlight a critical interplay between environmental factors and halo mass in shaping galaxy properties, affirming the environment as a fundamental driver in galaxy formation and evolution.

\end{abstract}



\section{Introduction}\label{sec:intro}
The size and mass of galaxies are fundamental properties that provide crucial insights into their formation and evolution. The galaxy size-mass relation (SMR), has been extensively studied to better understand the underlying processes that shape these properties \citep{Shen2003,Hyde2009,Bernardi2011a,Mosleh2013,Huang2017,Yoon2017,Rodriguez2021}. Significant work has been devoted to further understand the SMR and its dependence on intrinsic galactic properties (e.g. mass, concentration, morphology, star formation rate) and redshift \citep{Desroches2007,Hyde2009,Bernardi2011a,Bernardi2011b,vanderWel2014,Lange2015,Mowla2019,Rohr2022}. 

Other works focus on studying how galactic environment affects the SMR \citep{Cooper2012,Cappellari2013,Lani2013,Cebrian2014,Delaye2014,Shankar2014,Kelkar2015,Zhang2019,Yoon2017,Yoon2023, Ghosh2024}. Specifically, \cite{Yoon2017} show that the most massive early-type galaxies tend to be larger in dense environments, likely as a consequence of the numerous mergers they experience throughout their lifetimes. More recently, \cite{Ghosh2024} used a large sample of galaxies from the Hyper Suprime-Cam (HSC) survey to show a similar size-environment correlation at intermediate stellar masses, further reinforcing the idea that environmental influences extend across different mass regimes. This is in line with the idea that the size of massive dispersion-supported systems may be inflated over time if mergers impart sufficient energy \citep{BK2006,Kaufmann2007,Trujillo2011}. 

In contrast, low-mass galaxies, with their shallower gravitational potentials, are more susceptible to processes that remove or redistribute baryons \citep{Stinson2009,Muratov2015,ElBadry2016,Christensen2016}. Stellar feedback, such as supernova-driven outflows, can significantly alter their structure \citep{Ogiya2011,Pontzen2012,DiCintio2014,Onorbe2015,Chan2015,Lazar2020,Mercado2024}. Environmental mechanisms like ram pressure and tidal stripping further modulate the sizes of these low-mass galaxies, making them particularly vulnerable to external factors \citep{Haines2007,Peng2010,Wetzel2013,Moutard2018}. This implies that distinct astrophysical processes shape galaxy sizes at different mass scales: mergers and energetic encounters govern size growth at the high-mass end, while feedback and environmental stripping are especially important at lower masses.

A number of studies investigate the SMR for low- and intermediate-mass galaxies \citep[\mstar\ $\leq 10^{10}$ \msun;][]{Mosleh2013,Cebrian2014,Lange2015,Zhang2019,Kawinwanichakij2021,Carlsten2021,Rohr2022,Yoon2023,Klein2024}. In particular, \citet{Yoon2023} use Sloan Digital Sky Survey (SDSS) data to show that the SMR slope in this mass regime is shallower than in the high-mass regime. They also find that low-mass quenched galaxies -- which reside in more isolated environments -- are, on average, smaller than their counterparts in relatively denser regions. \cite{Benavides2023} also suggests that size evolution induced by tidal effects could be responsible for the existence of ultra diffuse galaxies (UDGs) at the low-mass end.

Despite the significant role of environment in shaping galaxy sizes, halo mass is still considered the primary driver of galactic structure, with environment playing a secondary role \citep[e.g.,][]{Kauffmann2004,Zheng2007}. Halo mass sets the depth of a galaxy’s gravitational well and influences the overall distribution of both baryons and dark matter, making it a fundamental determinant of a galaxy’s structural properties.

Another property closely linked to galaxy formation and evolution is the content of dark matter within the galaxy \citep{Dutton2010,Auger2010,Romeo2020,Kravtsov2024,Benavides2024}. The stellar-to-halo mass ratio (\mstar/\mhalo) provides key insight into the efficiency of star formation across different galaxy populations, with several studies exploring how it evolves with morphology, mass, and environment \citep[e.g.,][]{Romeo2020}. In particular, \cite{Kravtsov2024} shows that Milky Way satellites and some UDGs follow a relation between the dark matter content within a galaxy’s half-mass radius (\mdmhalf) and its stellar mass. However, large-scale simulations often struggle to reproduce these relations consistently, highlighting challenges in accurately modeling the galaxy-halo connection \citep{Romeo2020}. Understanding how the SMR and the \mdmhalf\ - \mstar\ relation depend on galactic environment for a broad range of low-mass galaxies is especially important as observatories like the Vera C. Rubin Observatory and Nancy Grace Roman Space Telescope usher in a new era of astronomy and give us access to statistical samples of low-mass galaxies across all environments.

In this paper, we employ a cosmological volume simulation to explore the effect that galactic environment has on the \rhalf\ - \mstar\ relation, the \mdmhalf\ - \mstar\ relation, and the \mhalo\ - \mstar\ relation. In particular, we aim to disentangle the effects of halo mass and environment on these scaling relations. We introduce our simulation and methodology in \S \ref{sec:sims}. In \S \ref{sec:results} we explore the dependence of the aforementioned scaling relations on environment. Finally, we summarize our results and discuss their implications in \S \ref{sec:conclusion}.

\section{Methodology}\label{sec:sims}

\subsection{The FIREbox Simulation Suite}\label{sec:firebox}
In this work we analyze simulated galaxies from the FIREbox cosmological volume simulation \citep{Feldmann2022}. While FIREbox is part of the Feedback In Realistic Environments (FIRE) project\footnote{\href{https://fire.northwestern.edu/}{https://fire.northwestern.edu/}} \citep{Hopkins2014,Hopkins2018,Hopkins2022}, it does not rely on the zoom-in method utilized by previous FIRE simulations \citep{Onorbe2014}. Rather, FIREbox simulates galaxy formation in a cubic cosmological volume $\sim (22.1 \, \rm cMpc)^3$ with periodic boundary conditions. Its initial conditions were generated with the MUltiscale Scale Initial Conditions \citep[MUSIC;][]{Hahn2011}, using cosmological parameters consistent with the Planck 2015 results \citep{Planck2016}: $\Omega_m = 0.3089$, $\Omega_{\Lambda} = 1 - \Omega_m$, $\Omega_b = 0.0486$, $h = 0.6774$, $\sigma_8 =  0.8159$, and $n_s = 0.9667$. A transfer function was calculated using the Code for Anisotropies in the Microwave Background \citep[CAMB;][]{Lewis2000,Lewis2011}.

FIREbox is run with the \textsc{Gizmo} code using the meshless-finite-mass method introduced by \cite{Hopkins2015}, as well as the FIRE-2 implementation to simulate gas cooling/heating, star formation, and stellar feedback \citep{Hopkins2018}. Star formation occurs in dense, molecular, self-gravitating and self-shielding gas, such that 100 per cent of the gas particles that meet these conditions are converted to stars. Specifically, the star formation occurs in gas at densities exceeding $300 \, \rm cm^{-3}$. Feedback includes energy, momentum, mass, and metal injected from supernovae (Type Ia and II), stellar winds from OB and AGB stars, radiative photo-ionization and photo-electric heating, and radiation pressure from young stars.

For the purposes of our analysis, we focus on \texttt{FB1024}, the primary simulation discussed in \cite{Feldmann2022}. \texttt{FB1024} contains 1024$^3$ baryonic and 1024$^3$ dark matter particles at the starting redshift ($z \, =$ 120) with masses of $m_{\rm b} = 6.3 \times 10^4$ \msun\ and $m_{\rm DM} = 3.3 \times 10^5$ \msun, respectively. The force-softening lengths for star and dark matter particles are $h_{\star} = 12 \, \rm pc$ and $h_{\rm DM} = 80 \, \rm pc$, while the force softening for gas particles is set to their smoothing length with a minimum value of 1.5 pc.

While FIREbox has a modest cosmological volume compared to some larger-scale hydrodynamical simulations \citep{Schaye2015,Crain2015,Pillepich2018,Nelson2018,Springel}, it still spans a wide range of galactic environments, from low-density voids to galaxy groups. The simulation contains halos with virial masses up to $\sim 10^{13}$ \msun, capturing environments comparable to small galaxy groups, though it does not fully sample the most massive cluster environments. Nevertheless, FIREbox offers an important advantage over larger-volume simulations: its ability to resolve the interstellar medium (ISM) at high spatial resolution. This allows for a more realistic treatment of star formation and feedback processes, which are crucial for understanding the structural and dynamical properties of low-mass galaxies. Given our focus on low-mass galaxies and the importance of resolving their internal baryonic processes, FIREbox is well-suited for investigating how environment influences galaxy sizes and dark matter content.

\subsection{Halo finding}\label{sec:halo Finding}

We identify and characterize dark matter halo properties with the AMIGA Halo Finder\footnote{\href{http://popia.ft.uam.es/AHF/}{http://popia.ft.uam.es/AHF/}} \citep[AHF;][]{Knollmann2009}. We define dark matter halos to be spherical systems with viral radii, \rvir, inside which the average density is equal to $\Delta(z) \rho_{\rm m}(z)$. The virial mass is then defined as the total mass within \rvir,
\begin{equation}
    M_{\rm vir} = (4/3)\pi \Delta(z) \rho_{\rm m}(z) r_{\rm vir}^3.
\end{equation}
Here, $\rho_{\rm m}(z)$ is the matter density at a given redshift and $\Delta(z)$ is the redshift-evolving virial overdensity defined by \citet{Bryan1998}. 

AHF identifies groups of galaxies, classifying the promary halos as ``centrals,'' and any subhalos orbiting within them as ``satellites.'' In addition, AHF truncates subhalos at their turnaround radii, identified near the local maxima of their radial density profiles, to more accurately represent the subhalo’s current bound structure. To maintain a consistent and physically motivated measure of halo size and mass for both centrals and satellites, we define the halo radius and mass for centrals to be their virial radius and mass, since they represent the full halo:
\begin{equation}
    r^{\rm centrals}_{\rm halo} \equiv r_{\rm vir}, \, \, {\rm and} \, \, M^{\rm centrals}_{\rm halo} \equiv M_{\rm vir}.
\end{equation}
For satellites, we take \rhalo\ to be the AHF-defined truncated radius ($r_{\rm t}$), reflecting the subhalo’s reduced extent due to tidal stripping and environmental effects, and \mhalo\ to be the total mass ($M_{\rm tot} = M_{\rm DM} + M_{\rm bary}$) within this truncated radius:
\begin{equation}
    r^{\rm satellites}_{\rm halo} \equiv r_{\rm t}, \, \, {\rm and} \, \, M^{\rm satellites}_{\rm halo} \equiv M_{\rm tot} (< \, r_{\rm t}).
\end{equation}

\subsection{Galaxy sample selection and definitions}\label{sec:sample and definitions}
In this analysis, we only consider halos that contain at least 128 gravitationally bound star particles at $z$ = 0. We define a galaxy's stellar mass to be the mass within the 3D spherical radius that contains 80 percent of the gravitationally bound stellar mass, following \citet{Mowla2019}:
\begin{equation}
    M_{\star} \equiv M_{\star}(< r_{\star}^{80}).
\end{equation}
We use the 3D spherical bound stellar half-mass radius, \rhalf, when discussing the size of a galaxy and denote the bound dark matter mass within this radius as \mdmhalf. We restrict our analysis to low-mass galaxies with \mstar\ $\leq 10^9$ \msun, which yields a sample size of N = 1218 halos (886 central and 332 satellite halos) with halo masses between $10^8$ \msun\ $\leq$ \mhalo\ $\leq 10^{11}$ \msun. This catalog was first identified by \citet{Moreno2022}, which visually recovers a small fraction of dark matter deficient galaxies missed by standard methods (see that work for details). Given that the processes influencing galaxy size are different at the high and low mass ends, we focus on the low-mass end in this paper. Additionally, FIREbox’s extensive and statistically robust sample of low-mass galaxies provides the necessary data to effectively investigate these processes.

\subsection{Quantifying environment}\label{sec:quantifying environment}

A number of different indicators have historically been used in the literature to quantify a galaxy's environment. For example, \cite{Christensen2024} use the distance to the nearest massive (\mvir\ $> 10^{11.5}$ \msun) galaxy to probe the environment of objects in their simulated sample. Others adopt the local number density of nearby neighbors as an estimator for environment \citep{Casertano1985,Mateus2004,Verley2007,Bluck2016}. Other works opt for more physically motivated quantities like the Tidal Index to estimate the magnitude of the tidal force an object feels due to its nearby massive neighbors \citep{Karachentsev1999,Karachentsev2013,Mutlu-Pakdil2024}. Finally, some work attempts to map out large-scale density fields using more sophisticated frameworks \citep{RomanoDiaz2007,Elek2022}. For a more detailed discussion of methods that quantify galactic environment, see \citet{Muldrew2012} and the references therein. 

\begin{figure*}
	\includegraphics[width=1.0\textwidth, trim = 0 0 0 0]{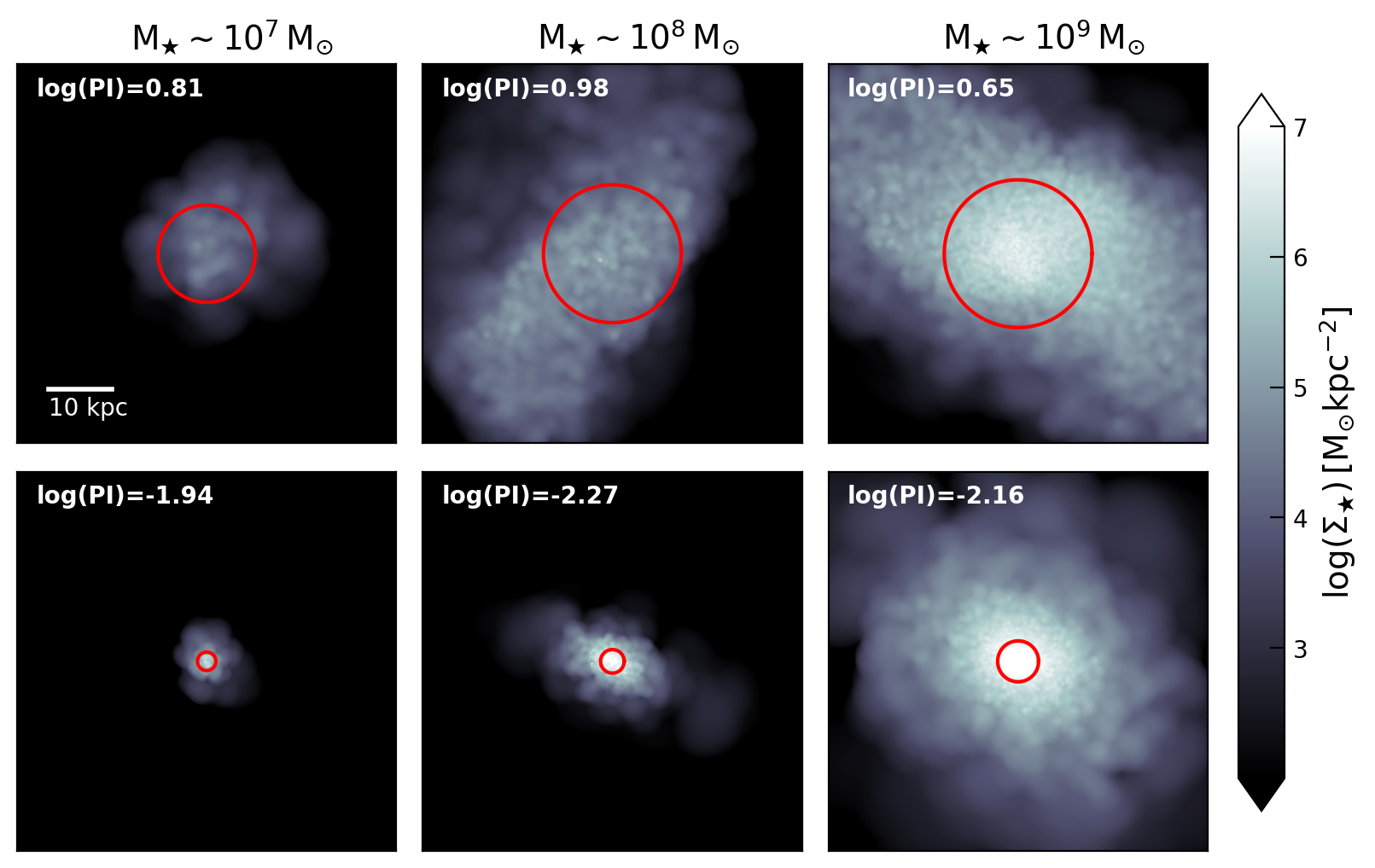}
	\centering
	    \caption{\textit{Stellar surface density maps of six representative galaxies.} We chose 3 different representative stellar masses (\textbf{left column:} \mstar\ $\sim \, 10^7$ \msun, \textbf{middle column:} \mstar\ $\sim \, 10^8$ \msun, \textbf{right column:} \mstar\ $\sim \, 10^9$ \msun). The red circles represent the 3D stellar half mass radius, \rhalf, for each galaxy. The galaxies displayed in the top row represent the galaxies with the largest \rhalf\ in its given mass bin, while the bottom row shows the galaxies with the smallest \rhalf. The logarithm of the PI value of each galaxy is reported at the top left of each panel. \textit{Takeaway:} Our simulated galaxies exhibit a wide range of sizes for fixed stellar masses. The Perturbation Index, a measure of the galaxy's environment, may play a role in bringing about this size variation.}
     	\label{fig:mocks}
\end{figure*}

For the purposes of this paper, we adopt the Perturbation Index (PI) to quantify galaxy environment \citep{Dahari1984,Verley2007,Choi2018,Martin2019,Jackson2021}. The PI offers several advantages over other environmental indicators. Specifically, it not only directly measures the tidal influence of nearby companions by accounting for both their mass and proximity but also incorporates the properties of the primary galaxy, such as its mass and size. This provides a more comprehensive assessment of environmental impact. Additionally, the PI’s ability to quantify the strength of tidal interactions provides a direct link to the physical processes (e.g., tidal stripping and mergers) that can significantly alter a galaxy’s size and dark matter distribution.

The PI quantifies the relative effect on a primary galaxy $p$ due to the tidal field around it and is defined by the ratio of the tidal force that galaxy feels (due to a companion $c$; $F^{c}_{\rm tidal}$) to its own binding force ($F^p_{\rm bind}$),
\begin{equation}
    {\rm PI} \equiv \sum \frac{F^c_{\rm tidal}}{F^p_{\rm bind}} \simeq \sum^{k\, (< r_{\rm max})}_{c=0}\left(\frac{M^c_{\rm halo}}{M^p_{\rm halo}} \right) \left(\frac{r^p_{\rm halo}}{D_{cp}} \right)^3,
\end{equation}
where $M^p_{\rm halo}$ and $r^p_{\rm halo}$ are the halo mass and radius of the primary galaxy, $M^c_{\rm halo}$ is the halo mass of a companion galaxy $c$, and $D_{cp}$ is the 3D distance from the primary to its companion galaxy. This formulation of the PI estimates the cumulative tidal field about a galaxy due to the nearest $k$ neighbors out to a maximum radius, $r_{\rm max}$. We calculate PI for each galaxy in our sample, summing contributions from all galaxies out to a maximum radius of $r_{\rm max} = \sqrt{3} L_{\rm box}/2$ to account for interactions throughout the full simulation volume while handling periodic boundary conditions.

We use halo masses and radii in our formulation, rather than the stellar mass and galactic sizes (\rhalf), given that tidal effects are certainly important out to \rhalo. This choice results in a PI distribution that is more naturally centered around unity, such that a PI $\gg$ 1 represents a galaxy that is highly perturbed by its surroundings (i.e. $F_{\rm tidal} > F_{\rm bind}$).

To assess the physical scale that PI probes, we compare PI values calculated using smaller values of  $r_{\rm max}$  and find that for most galaxies, PI remains unchanged when restricting to  $r_{\rm max}$ = 10  Mpc , but decreases when limiting to  $r_{\rm max}$ = 1  Mpc. This suggests that PI is primarily sensitive to tidal interactions on scales of 1–10 Mpc rather than sub-Mpc interactions. Consequently, while PI effectively captures local tidal effects, it does not directly probe large-scale cosmic structures, which have been shown to influence galaxy properties in other studies \citep[e.g.,][]{Muldrew2012}.

Figure \ref{fig:mocks} displays the projected stellar surface mass density maps of six example galaxies at three representative stellar masses (left-to-right columns: \mstar\ $\sim \, 10^7$ \msun, $\sim \, 10^8$ \msun, and  $\sim \, 10^9$ \msun, respectively). The red circles represent each galaxy's 3D stellar half mass radius (\rhalf). The top (bottom) row illustrates the largest (smallest) galaxies in each respective stellar mass bin. We also report the logarithm of each galaxy's PI value at the top left of each panel. While many galaxies in our sample may have similar stellar masses, their sizes may vary significantly. For these examples, the more extended galaxies in each stellar mass bin have larger PI values, hinting at a possible role of environment in setting a galaxy's relative extent (more below).


\subsection{Random forest analysis}\label{sec:random forest}

Given the inherently complex nature of galaxy formation and evolution, it is plausible that a number of processes influence a galaxy’s size and inner dark matter content. To better capture the intricacy of these relationships, we adopt a machine learning approach that allows us to explore our data without confining the analysis to a search for simple power-laws. Specifically, we utilize a Random Forest (RF) regressor \citep{Ho1995}. 

An RF regressor can directly model nonlinear relationships between input parameters (commonly referred to as ``features'') and the output quantities of interest (known as ``target variables''). Features are the independent variables used as inputs to the model, while target variables represent the dependent quantities the model is trained to predict. In our case, the features correspond to global galactic and environmental properties, while the target variables quantify deviations in galaxy sizes and inner dark matter content from their expected median relations. 

An important advantage of this approach is that Random Forest regression not only predicts these outcomes but also provides a measure of feature importance, allowing us to identify which properties most significantly influence a galaxy’s structural and dark matter characteristics.

In this analysis, we focus on three features that are physically relevant to galaxy formation and structure: the halo mass (\mhalo), the stellar-to-halo mass ratio (\mstar/\mhalo), and the Perturbation Index (PI). Halo mass sets the galaxy’s gravitational framework and regulates the distribution of baryons and dark matter on a cosmological scale. The stellar-to-halo mass ratio provides insight into star formation efficiency over a galaxy’s history. The Perturbation Index encodes environmental factors, such as tidal forces and interactions, that can reshape a galaxy’s structure. By training our model on $\log$(\mhalo), $\log$(\mstar/\mhalo), and $\log$(PI), we aim to isolate the most influential drivers behind deviations in galaxy sizes and inner dark matter content from median scaling relations.

\subsubsection{Overview}\label{sec:rf overview}
Random Forest regression is an ensemble learning method that constructs a multitude of decision trees during training and outputs the average prediction of the individual trees to improve predictive accuracy and reduce variance \citep{Breiman2001}. 

Each tree consists of a series of binary decision splits, where at each node, the algorithm selects a feature and a corresponding threshold (splitting criterion) that best separates the data. In our implementation, we use mean squared error (MSE) as the splitting criterion to determine the optimal split at each node. MSE measures the average squared difference between the true and predicted values, ensuring that larger errors contribute more significantly to the loss function.

Formally, MSE is computed as:
\begin{equation}
\text{MSE} = \frac{1}{N} \sum_{i=1}^{N} (y_i - \hat{y}_i)^2,
\end{equation}
where $y_i$ are the target values, $\hat{y}_i$ are the model predictions, and N is the total number of samples.

The splitting process continues recursively until a stopping condition is met, such as reaching a predefined maximum tree depth or a minimum number of samples per leaf node. A leaf node is the final, unsplit node of the tree, which contains the predicted output value. 

Each decision tree is trained on a bootstrap sample, meaning it is built using a randomly drawn subset of the training data (with replacement). This bootstrap aggregation (or “bagging”) ensures that each tree is trained on a slightly different dataset, increasing robustness and reducing sensitivity to individual data points. Additionally, at each split, the algorithm only considers a randomly selected subset of available features, further promoting variation among the trees and improving model generalization.

While the primary function of a Random Forest regressor is to model the relationship between input features and target variables, our main objective in this study is to utilize it as a tool to assess the relative importance of various galactic and environmental features in predicting deviations from median scaling relations. Given our relatively small sample size (N $\sim$ 1200), achieving high predictive accuracy is inherently challenging. The limited dataset may not fully capture the statistical diversity of the physical processes at play, potentially restricting the model’s ability to generalize. 

Thus, instead of prioritizing predictive performance, we focus on feature importance. To quantify this, we use  $R^2$  (coefficient of determination), defined as:
\begin{equation}\label{equation:r_sq}
R^2 = 1 - \frac{\sum_i (y_i - \hat{y}_i)^2}{\sum_i (y_i - \bar{y})^2},
\end{equation}
where $\bar{y}$ is the mean of the true target values.  A higher  $R^2$  indicates that more variance in the target variable is explained by the model, but in this analysis, it primarily serves as a reference rather than an optimization goal. By emphasizing the identification of influential features, we aim to gain insights into the underlying physical processes that influence galaxy properties, despite the constraints imposed by the sample size.

\subsubsection{Hyperparameters, data preparation and model training}\label{sec:Data Prep}
The RF algorithm has several important hyperparameters that control the complexity of the model:
\begin{itemize}
    \item \texttt{n\_estimators}: The number of trees in the forest. A larger number typically yields more stable predictions but increases computational cost. This value also saturates, so there is always a minimum number of trees for a given size of the dataset that will lead to stable predictions
    \item \texttt{max\_depth}: The maximum depth of each tree. Limiting depth can prevent over-fitting by stopping splits once trees become too complex.
    \item \texttt{min\_samples\_leaf}: The minimum number of samples required at a leaf node. Increasing this value leads to simpler trees that generalize better.
    \item \texttt{max\_features}: The number (or fraction) of features to consider at each split. Controlling this affects the diversity of trees in the ensemble and the degree to which each tree is required to explore the full set of features, rather than the best performing one..
    \item \texttt{min\_samples\_split}: The minimum number of samples required for an internal node to be split. Increasing this value reduces the likelihood of creating deep, highly specialized trees by requiring more data to justify a split, thereby promoting broader, more generalized decision boundaries.
\end{itemize}
Before training the model, we preprocess our features using RobustScaler from the \textsc{scikit-learn} \citep{Pedregosa2011}. This scaler removes the median and scales the data according to the interquartile range (IQR), making it robust to outliers. This scaling centers the features around their median and scales them based on the 25th and 75th percentiles, making the scaling robust to extreme values. 

We identify the optimal hyperparameters using GridSearchCV with five-fold cross-validation. Rather than optimizing the entire hyperparameter space, we focus solely on optimizing \texttt{max\_features} and \texttt{min\_samples\_leaf} to maximize $R^2$. We fix \texttt{n\_estimators} at 600 to ensure stable predictions without excessive computational cost and leave all other parameters at their default values to maintain simplicity. By concentrating on optimizing \texttt{max\_features} and \texttt{min\_samples\_leaf}, we directly control tree complexity directly, tuning only the parameters most likely to influence our model’s bias-variance trade-off given our small dataset. This approach reduces the risk of over-fitting to our limited sample while still enabling some meaningful optimization of model complexity.

Although we use MSE as the RF criterion, we use  $R^2$  in GridSearchCV because it provides a more intuitive measure of model performance. The two metrics are mathematically related: minimizing MSE maximizes  $R^2$, since the numerator of the second term in Equation \ref{equation:r_sq}  is simply the total MSE. Additionally,  $R^2$  is unitless, allowing for easier comparison across different target variables.

After selecting the optimal hyperparameters, we apply them in all subsequent analyses. To ensure the robustness of our findings and that results are not dependent on any single train-test split, we repeat the random train-test splitting process 500 times, each time training the Random Forest regressor on 50\% of the data and testing on the other 50\%. This procedure yields distributions of performance metrics and feature importance scores, allowing us to assess their stability and variability.

\subsubsection{Feature importance analysis}\label{sec:Importance analysis}
Feature importance in Random Forest regression provides a quantitative measure of the contribution of each feature to the model’s predictions.The importance of a feature is computed as the total weighted reduction in MSE across all nodes where the feature is used for splitting.

Mathematically, for a feature  $x_j$ , its importance score is given by:
\begin{equation}
I(x_j) = \sum_{\text{splits on } x_j} \frac{N_{\text{node}}}{N_{\text{total}}} \Delta \text{MSE},
\end{equation}
where  $N_{\text{node}}$  is the number of samples at the node where the split occurs, and  $\Delta$ MSE  is the decrease in mean squared error due to the split.

After training the model in each of the 500 iterations, we extract the feature importance scores, producing distributions of importance scores for each feature.  This method offers a comprehensive understanding of which properties -- halo mass, stellar-to-halo mass ratio, or environment -- dominate in shaping deviations in galaxy sizes and inner dark matter content from the median scaling relations.

\section{Results}\label{sec:results}

\subsection{Scaling relations in FIREbox}\label{sec:scaling relations}

\begin{figure*}
	\includegraphics[width=1.0\textwidth, trim = 0 0 0 0]{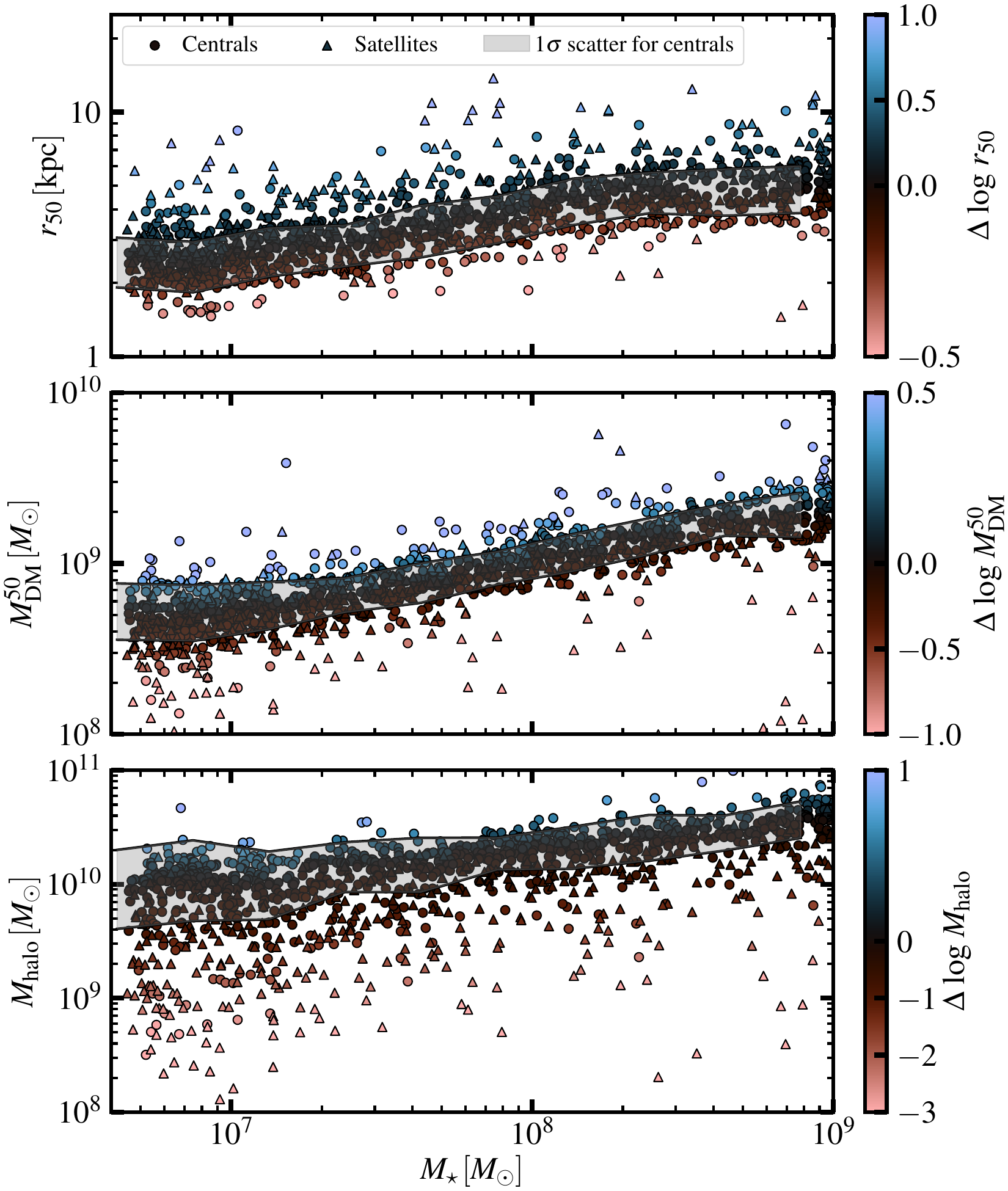}
	\centering
	    \caption{\textit{The \rhalf\ - \mstar\, \mdmhalf\ - \mstar\ and \mhalo\ - \mstar\ relations.} The stellar half mass radius (\rhalf; \textbf{top panel}), the dark matter mass within \rhalf\ (\mdmhalf; \textbf{middle panel}), and the total halo mass (\mhalo; \textbf{bottom panel}) versus the stellar mass, \mstar, for our FIREbox galaxy sample with \mstar\ $\leq \, 10^9$ \msun. Filled-in circles represent centrals, while the satellites appear as filled-in triangles. The gray, shaded region represents one standard deviation above and below the median behavior of the centrals. All points are color-coded by the residuals about the median behavior of central galaxies for both relations. \textit{Takeaway:} The residuals (\drhalf, \ddm, and \dmhalo) quantify the relative extent and relative dark matter content of a galaxy compared to centrals of similar stellar mass.}
     	\label{fig:scaling_relations}
\end{figure*}

Figure \ref{fig:scaling_relations} presents the \rhalf\ - \mstar\ (top panel), \mdmhalf\ - \mstar\ (middle panel), and \mhalo\ - \mstar\ relations (bottom panel) for FIREbox-simulated galaxies with \mstar\ $\leq \, 10^9$ \msun\footnote{We note that four galaxies have stellar masses below $4 \times 10^6$ \msun, but we exclude them from this particular plot to maintain a clearer presentation. However, we still include these galaxies in the rest of our analysis}. We display central halos as filled-in circles, while the filled-in triangles represent satellites. The gray shaded region represents one standard deviation above and below the median behavior of the central halos. We color-code points by the residuals, or logarithmic distance from the median behavior of central galaxies for all three relations (denoted as \drhalf, \ddm, and \dmhalo, respectively). We use these residuals to quantify how extended/compact, DM rich/poor, or DM over-massive/under-massive a galaxy is compared to other central galaxies of similar stellar mass. In other words, we consider galaxies with positive residuals to be extended, DM rich, and DM over-massive while galaxies with negative residuals are compact, DM poor, or DM under-massive.

We note that the asymmetric ranges of the colorbars can be slightly misleading. In the top panel, for instance, the galaxies shown by the palest red symbols are not as far from the centrals’ median behavior as those depicted by the palest blue symbols. Conversely, in the middle and bottom panels, the palest red symbols represent galaxies that lie farther from their respective medians than the palest blue symbols.

\begin{figure}
	\includegraphics[width=\columnwidth, trim = 0 0 0 0]{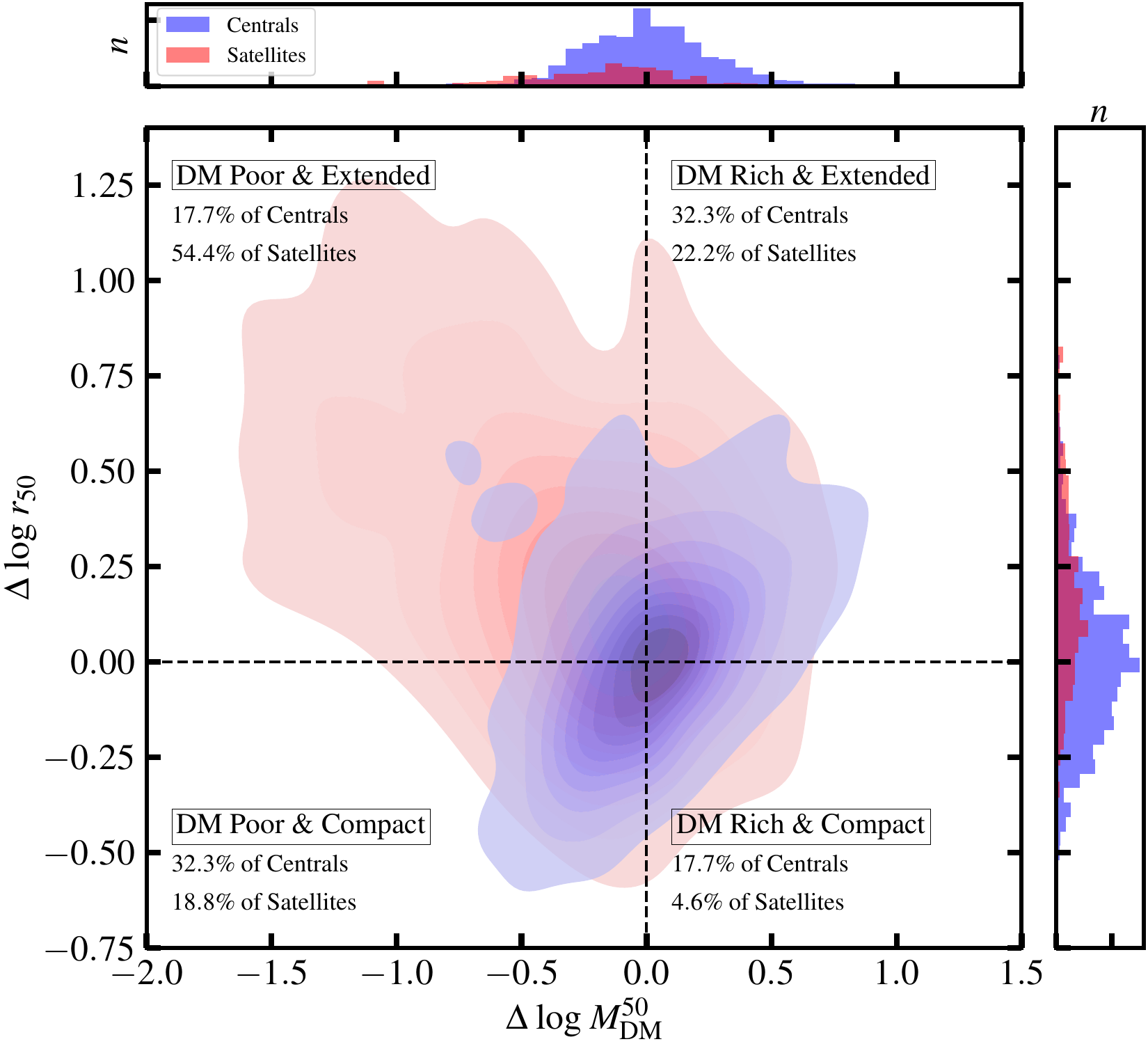}
	\centering
    	\caption{\textit{A 2D kernel density estimate of galaxy relative extent versus relative inner dark matter content.} \drhalf\ versus \ddm\ for centrals (blue) and satellites (red) in our sample. The space is divided into four quadrants that separate our sample into four subpopulations: DM rich and extended (\textbf{Quadrant I}), DM poor and extended (\textbf{Quadrant II}), DM poor and compact (\textbf{Quadrant III}), and DM rich and compact (\textbf{Quadrant IV}). The joint plots with shared axes display the 1D histograms that show the behavior of the central and satellite populations. \textit{Takeaway:} More than half of the satellites in our sample are DM poor and extended, while the centrals appear concentrated closer to the origin in this space -- pointing to a possible role of environment in determining a galaxy's relative extent and relative inner dark matter content.}
     	\label{fig:kde}
\end{figure}

Figure \ref{fig:kde} explores the relationship between relative extent and relative inner dark matter content (\drhalf\ versus \ddm) for our sample with a 2D kernel density estimation for centrals (blue) and satellites (red). We divide this space into four quadrants that separate our sample into four subpopulations: DM rich and extended (Quadrant I), DM poor and extended (Quadrant II), DM poor and compact (Quadrant III), and DM rich and compact (Quadrant IV). The joint panels with shared axes display the 1D histograms that illustrate the behavior of the central and satellite populations. More than half of the satellite galaxies in our sample are dark-matter poor and relatively extended, a pattern consistent with the dark matter depletion and tidal stripping they likely undergo while orbiting their host halos \citep{Moore1999,Mayer2001,Kravtsov2004,Penarrubia2008,Donghia2009,Kazantzidis2011,Tomozeiu2016,Frings2017,Fattahi2018}. On the other hand, central galaxies cluster near the origin of this parameter space, tracing a diagonal that indicates a slight positive correlation between their relative extent and relative inner DM content. This contrast in the distributions of centrals and satellites suggests that the galactic environment likely plays a role in determining a galaxy’s position on the \rhalf\ – \mstar\ and \mdmhalf\ – \mstar\ relations.

We note that seven DM poor and compact galaxies (in Quadrant III) have \ddm\ values smaller than -2. These galaxies are the same seven dark matter deficient galaxies highlighted by \citet{Moreno2022}. Because they are outliers formed through a highly unusual mechanism, we exclude them from the remainder of our analysis.

\subsection{Environmental effects on scaling relations}\label{sec:environmental effect}

\begin{figure*}
	\includegraphics[width=1.0\textwidth, trim = 0 0 0 0]{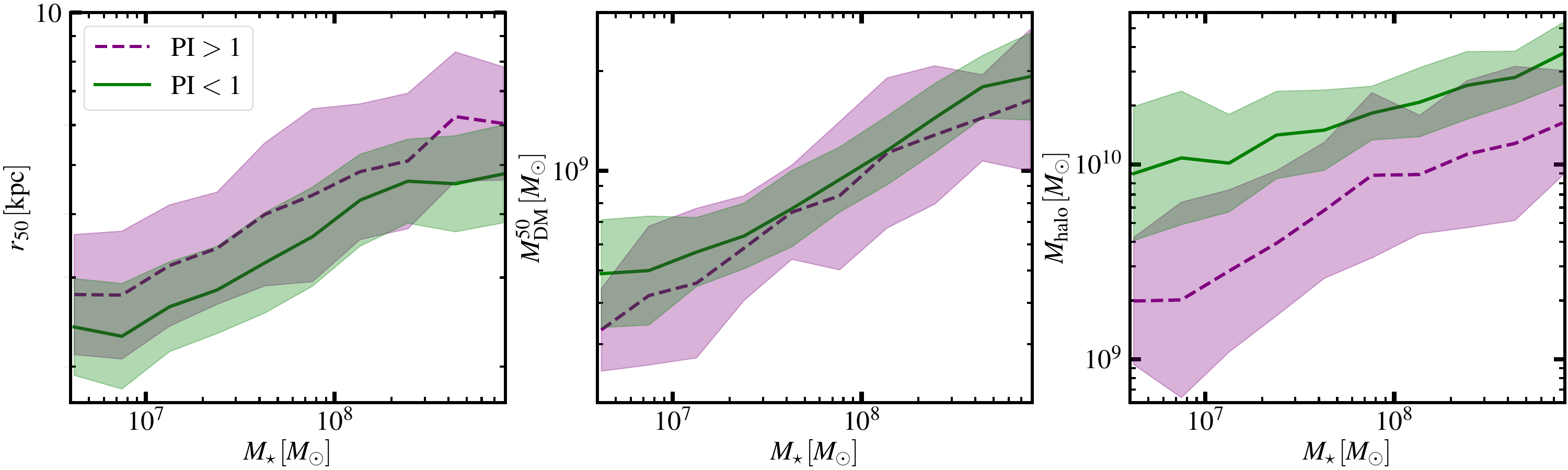}
	\centering
	    \caption{\textit{Environmental effect on the scaling relations.} The median behavior of the same three scaling relations introduced in Figure \ref{fig:scaling_relations} after splitting our sample into two subpopulations: galaxies with PI $>$ 1 (purple dashed lines) and galaxies with PI $<$ 1 (green solid lines). The purple and green shaded regions represent the 1$\sigma$ scatter about the median behavior of the two samples. \textit{Takeaway}: At fixed stellar mass, more-perturbed galaxies with larger PI values are, in general, more extended and have lower halo masses than their less-perturbed counterparts.}
     	\label{fig:env_effect_scaling_relations}
\end{figure*}

\begin{figure*}
	\includegraphics[width=1.0\textwidth, trim = 0 0 0 0]{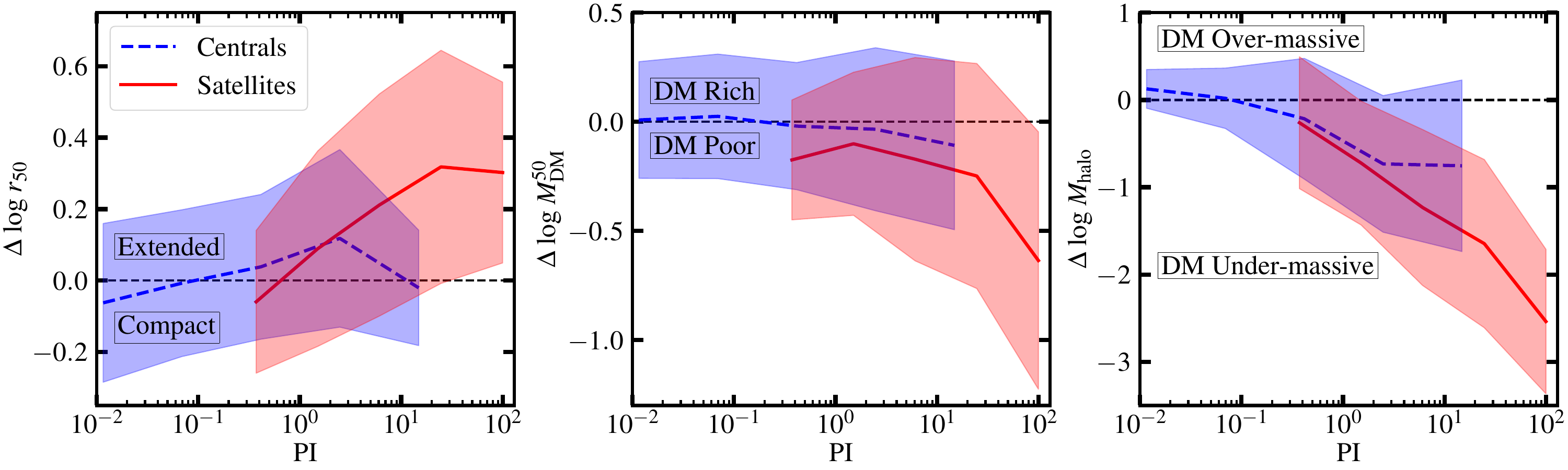}
	\centering
	    \caption{\textit{Environmental dependence on scaling relation residuals.} The relative extent (\drhalf; \textbf{left panel}), relative DM content (\ddm; \textbf{middle panel}), and relative halo massiveness (\dmhalo; \textbf{right panel}) as a function of PI. The blue, dashed lines track the median behavior of the centrals and the red, solid lines track the same for the satellites. The blue and red shaded regions display the $1\sigma$ deviation above and below the median behavior for both populations. The horizontal black dashed line marks the zero point on the vertical axis and represents the median behavior of centrals on the the three scaling relations introduced in Figure \ref{fig:scaling_relations}. \textit{Takeaway}: Both centrals and satellites display a relationship between \drhalf\ and \dmhalo, and PI. Specifically, galaxies with higher PI values tend to be extended and DM under-massive when compared to their counterparts with lower PIs.}
     	\label{fig:Delta_v_PI}
\end{figure*}

Figure \ref{fig:env_effect_scaling_relations} presents the scaling relations in Figure \ref{fig:scaling_relations} again, but displaying the median behavior of galaxies with PI $>$ 1 (purple dashed lines) or PI $<$ 1 (green solid lines). The purple and green shaded regions represent the 1$\sigma$ scatter about the median behavior of either population. 

We find that galaxies with larger PI values have their size–mass relation (SMR) systematically shifted toward larger sizes and their \mhalo\ – \mstar\ relation positioned at lower dark matter content compared to galaxies with smaller PI values. In other words, \textit{at fixed stellar mass}, galaxies more strongly perturbed by their environment tend to be more extended and dark matter under-massive than their less-perturbed counterparts. This difference likely arises from increased dark matter stripping and more frequent tidal interactions in more crowded environments \citep{BK2006,Kaufmann2007,Donghia2010,Trujillo2011,Yoon2017}. Interestingly, galaxies with both high and low PI values follow a similar \mdmhalf\ – \mstar\ relation, except perhaps at the lowest masses (\mstar\ $< \, 10^7$ \msun), where the inner dark matter content is likely more sensitive to environmental influences and star formation feedback due to shallower potential wells \citep[see][]{ElBadry2016}. While it is clear that the environment contributes to the scatter in these scaling relations, it remains an open question whether there is a direct relationship between a galaxy’s relative extent or inner dark matter content and its surroundings.

Figure \ref{fig:Delta_v_PI} illustrates the extent to which environment (PI) sets how extended/compact, DM rich/poor, or DM over-massive/under-massive a galaxy can be at fixed stellar mass. We show the residuals \drhalf\ (left panel), \ddm\ (middle panel), and \dmhalo\ (right panel) as a function of PI. The dashed blue line and shaded region represent the median behavior of the centrals and the 1$\sigma$ scatter about the median while the solid red line and shaded region illustrates the same for the satellites in our sample. Finally, the horizontal, black dashed lines mark the zero-point on the $y$-axis and represent the median behavior of centrals on the three different scaling relations we study (see Figure \ref{fig:scaling_relations}). Values above (below) the horizontal line represent galaxies that are extended (compact), DM rich (poor), or DM over-massive (under-massive) in the left, middle, and right panels, respectively. 

Similarly to Figure \ref{fig:env_effect_scaling_relations}, we find that galaxies experiencing stronger environmental perturbations (i.e., those with higher PI values) tend to be more extended and have reduced dark matter content relative to less-perturbed galaxies of comparable stellar mass. This trend holds for both central and satellite galaxies with the highest PI values. However, the relationship is relatively weak, indicating that while a galaxy’s environment does influence its position on these scaling relations, it is not the sole factor responsible for the observed scatter. Furthermore, the negative correlation between \ddm\ and PI is only evident among the most perturbed satellites, which also happen to be the galaxies with the lowest stellar masses.

\subsection{Random forest regression}\label{sec:rf regressioin}

While Figure \ref{fig:Delta_v_PI} demonstrates that a galaxy’s position on these scaling relations depends, at least to some extent, on its environment (as quantified by PI), it is natural to ask which global properties most strongly influence how far above or below the median relations for centrals a galaxy resides. To investigate this, we turn to the Random Forest regression model introduced in Section \ref{sec:random forest}. Specifically, we train our model on logarithm of the following features: halo mass ($\log$ \mhalo), the stellar-to-halo mass ratio ($\log$ \mstar/\mhalo), and the Perturbation Index ($\log$ PI) to predict a galaxy’s relative extent (\drhalf) and its relative inner dark matter content (\ddm).

After training the Random Forest regressor using the optimal hyperparameters determined via GridSearchCV (as listed in Table \ref{tab:hyperparams_results}), we evaluate the model’s performance on the test sets across the 500 iterations. For the relative extent (\drhalf), the mean $R^2$ score on the test sets was approximately 0.32, indicating that the model explains approximately 32\% of the variance in the deviations of the half-mass radius from the median size–mass relation for centrals at fixed stellar mass. The corresponding mean MSE is approximately 0.1, providing a direct measure of the average squared error in the model’s predictions.

For the relative inner dark matter content (\ddm), the mean $R^2$ score was approximately 0.41, suggesting that the model accounts for about 41\% of the variance in the deviations of the dark matter mass within \rhalf\ from the median \mdmhalf\ – \mstar\ relation for centrals at fixed stellar mass. The mean MSE for \ddm\ is approximately 0.2, reflecting the typical magnitude of prediction errors. While these $R^2$ values indicate moderate predictive power, our primary focus is on the relative importance of the features rather than the overall predictive accuracy.

\begin{figure}
	\includegraphics[width=\columnwidth, trim = 0 0 0 0]{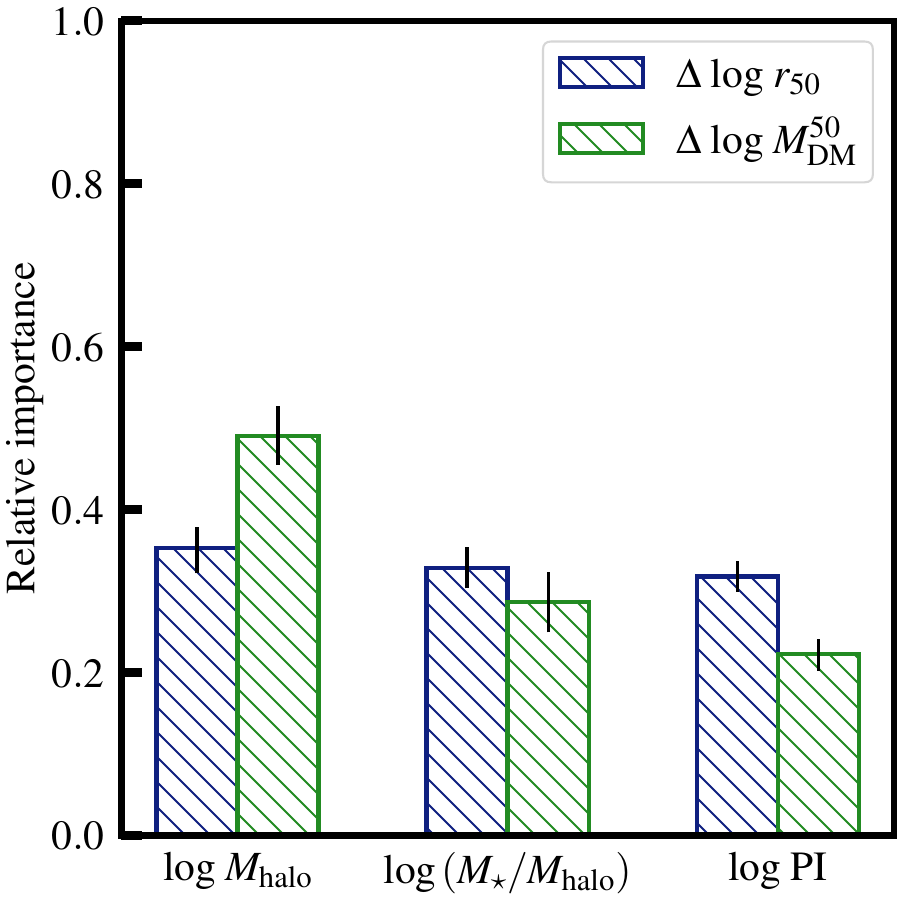}
	\centering
	    \caption{\textit{RF model feature importance.} The relative parameter importance in predicting relative extent (\drhalf; blue bars) and relative inner dark matter content (\ddm; green bars). Bar heights represent the median importance for each parameter, while the error bars mark the 5th and 95th percentiles of the importance distribution over the 500. We list the input features on the horizontal axis. \textit{Takeaway:} All three features are similarly as important in predicting \drhalf, while \mhalo\ is most significant in predicting \ddm.}
     	\label{fig:feature_importance}
\end{figure}

\begin{table}
    \centering
        \begin{tabular}{lcccc}
        \hline
        \textbf{Target} & \textbf{max} & \textbf{min} & \textbf{$\langle R^{2} \rangle$} & {\color{violet}\textbf{$\langle {\rm MSE} \rangle$}} \\
        \textbf{Variable} & \textbf{features} & \textbf{samples leaf} &  &    \\
        \hline
        \drhalf & `sqrt' & 3 & 0.32 & {\color{violet}0.10} \\
        \ddm    & `sqrt' & 3 & 0.41 & {\color{violet}0.20} \\
        \hline
        \end{tabular}
        \caption{\textit{Optimal hyperparameters and performance.} The table presents the optimized hyperparameters alongside the mean $R^2$ and MSE values for the two target variables (\drhalf\ and \ddm), averaged over 500 iterations.}
    \label{tab:hyperparams_results}
\end{table}

Figure \ref{fig:feature_importance} shows the relative parameter importance in predicting relative extent (\drhalf; blue bars) and relative inner dark matter content (\ddm; green bars), extracted from our Random Forest regressor model.\footnote{Setting \texttt{max\_features=None} instead of \texttt{max\_features=`sqrt’} alters the absolute importance values but does not change the relative ranking of features.} The bar heights represent the median importance for each parameter, while the error bars mark the 5th and 95th percentiles of the importance distribution created by repeating the experiment 500 times. The input features are listed on the horizontal axis.

We find that, when predicting a galaxy’s relative extent (\drhalf), all three properties (\mhalo, \mstar/\mhalo, PI) are similarly important. This suggests that what sets a galaxy’s relative size is a highly complex interplay between mass and environment. The significant importance of $\log$ \mhalo\ indicates that halo mass plays a crucial role in determining deviations in galaxy size from the median scaling relation. The importance of $\log$ PI highlights the influence of the environment, suggesting that environmental interactions can directly affect galactic sizes. The contribution of $\log$ \mstar/\mhalo\ implies that the efficiency of star formation also impacts the relative extent of galaxies.

On the other hand, the halo mass ($\log$ \mhalo) is the most important predictor of relative inner dark matter content (\ddm). The feature importance analysis reveals that $\log$ \mhalo\ has a significantly higher median importance score than the other features when predicting \ddm. This underscores the fundamental role of halo mass in shaping the dark matter distribution within galaxies. The other features exhibit lower importance scores, indicating that they have a lesser impact on predicting deviations in inner dark matter content from the median relation.

We note that the error bars, representing the 5th and 95th percentiles, are narrow relative to the height of each bar. This low variability in the importance scores suggests that our findings are robust against different train-test splits and that the identified feature importances are stable across the 500 iterations. These findings indicate that the environment, quantified by the Perturbation Index, and the halo mass both play significant roles in influencing galaxy properties. The influence of PI on \drhalf\ suggests that environmental mechanisms can directly affect galaxy sizes, possibly through mechanisms such as tidal interactions, mergers, or ram-pressure stripping. Additionally, since the halo mass is also affected by environmental factors at fixed stellar masses, as shown in Figure \ref{fig:env_effect_scaling_relations}, the environment may indirectly influence galaxy properties through its impact on halo mass assembly.

\begin{figure*}
	\includegraphics[width=1.0\textwidth, trim = 0 0 0 0]{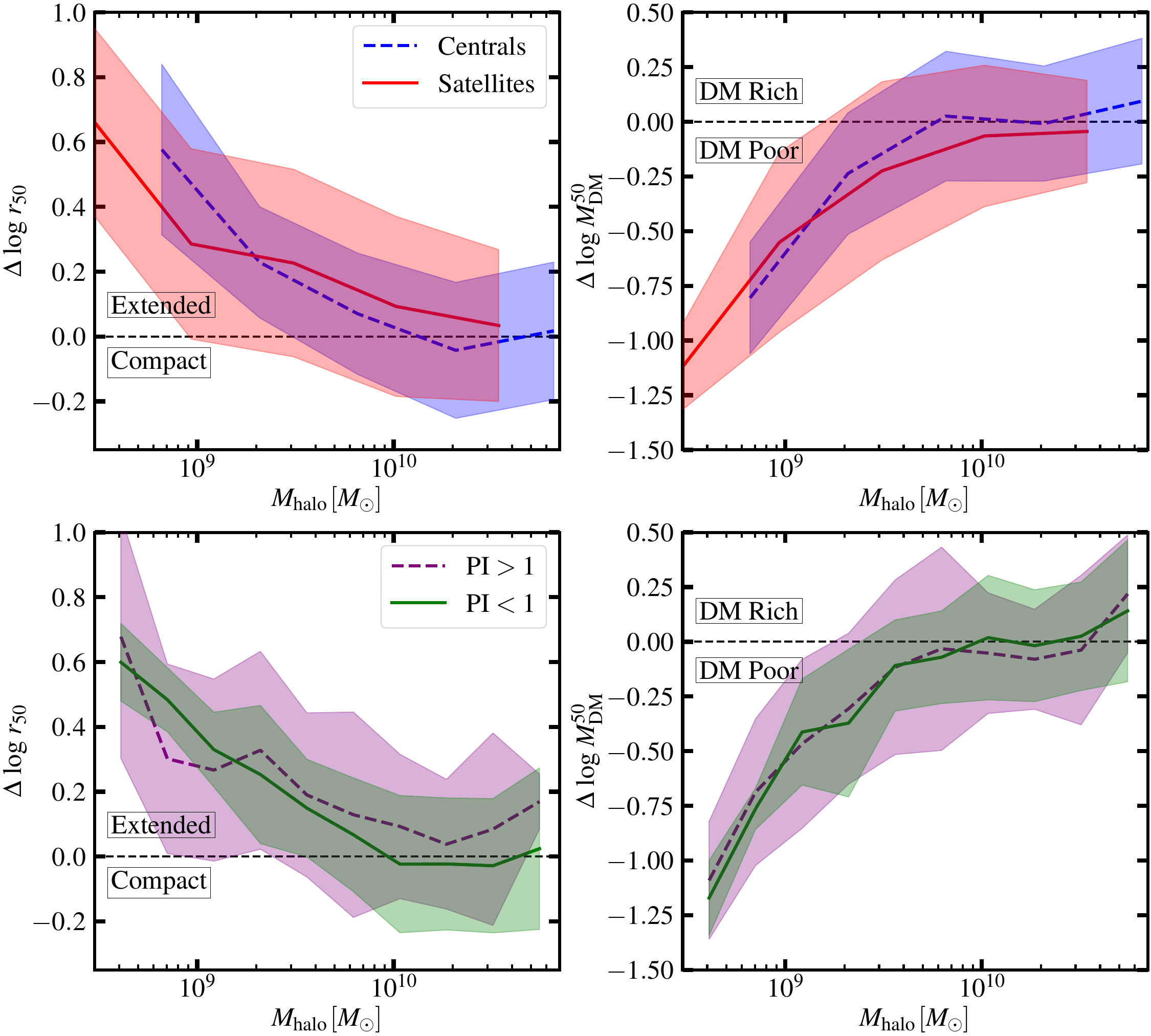}
	\centering
	    \caption{\textit{Relative extent and inner DM content versus \mvir.} \drhalf\ (\textbf{left panels}) and \ddm\ (\textbf{right panels}) as a function of halo mass.  The top and bottom panels illustrate the same relationship. However, in the \textbf{top panels}, we split our sample into centrals (blue, dashed lines) and satellites (red, solid lines), while the \textbf{bottom panels} show our sample split into galaxies with PI greater than (purple, dashed line) and less than (green, solid line) unity. The shaded regions illustrate the 1$\sigma$ scatter about the lines. \textit{Takeaway:} Regardless of how the sample is separated, galaxies with lower halo masses are more extended and more DM poor than galaxies at higher masses.} 
     \label{fig:Delta_v_mvir}
\end{figure*}

In summary, our Random Forest analysis reveals that a galaxy’s relative extent (\drhalf) is determined by a combination of halo mass, stellar-to-halo mass ratio, and environmental influences. This indicates that both intrinsic properties and external factors contribute to deviations in galaxy sizes from the median SMR for centrals. In contrast, the relative inner dark matter content (\ddm) is predominantly influenced by the halo mass, highlighting the fundamental role of halo mass in determining the dark matter distribution within the innermost galactic regions.

\subsection{Disentangling environmental and mass effects}\label{sec:env and mvir}

The results from our Random Forest analysis indicate a complex interplay between intrinsic properties (such as halo mass and stellar-to-halo mass ratio) and the environment. However, the degree to which environment directly shapes galaxy properties -- as opposed to indirectly doing so through its effect on halo mass at fixed stellar mass -- remains unclear. In the following section, we focus on carefully separating these influences to gain a more nuanced understanding of how mass and environment together sculpt the diversity of galaxy sizes and inner dark matter content.

Figure \ref{fig:Delta_v_mvir} reports the relative extent (\drhalf; left panels) and relative inner dark matter content (\ddm; right panels) as functions of halo mass (\mhalo). In the top panels, we divide our sample into centrals and satellites, with the blue and red dashed lines and regions tracing the median behavior and 1$\sigma$ scatter of these subpopulations, respectively. The  bottom panels present the same data, but now split by environmental perturbation: the purple dashed lines represent galaxies with PI $>$ 1, while the green solid lines represent those with PI $<$ 1. Regardless of how the sample is separated (by central/satellite status or by environmental perturbation), one robust trend emerges: \textit{galaxies with lower halo masses tend to be more extended and more DM poor than their higher-mass counterparts at fixed stellar mass.} This result, as alluded to by the Random Forest regression results, suggests that a galaxy’s relative extent and relative inner dark matter content depend strongly on halo mass.

\begin{figure*}
	\includegraphics[width=1.0\textwidth, trim = 0 0 0 0]{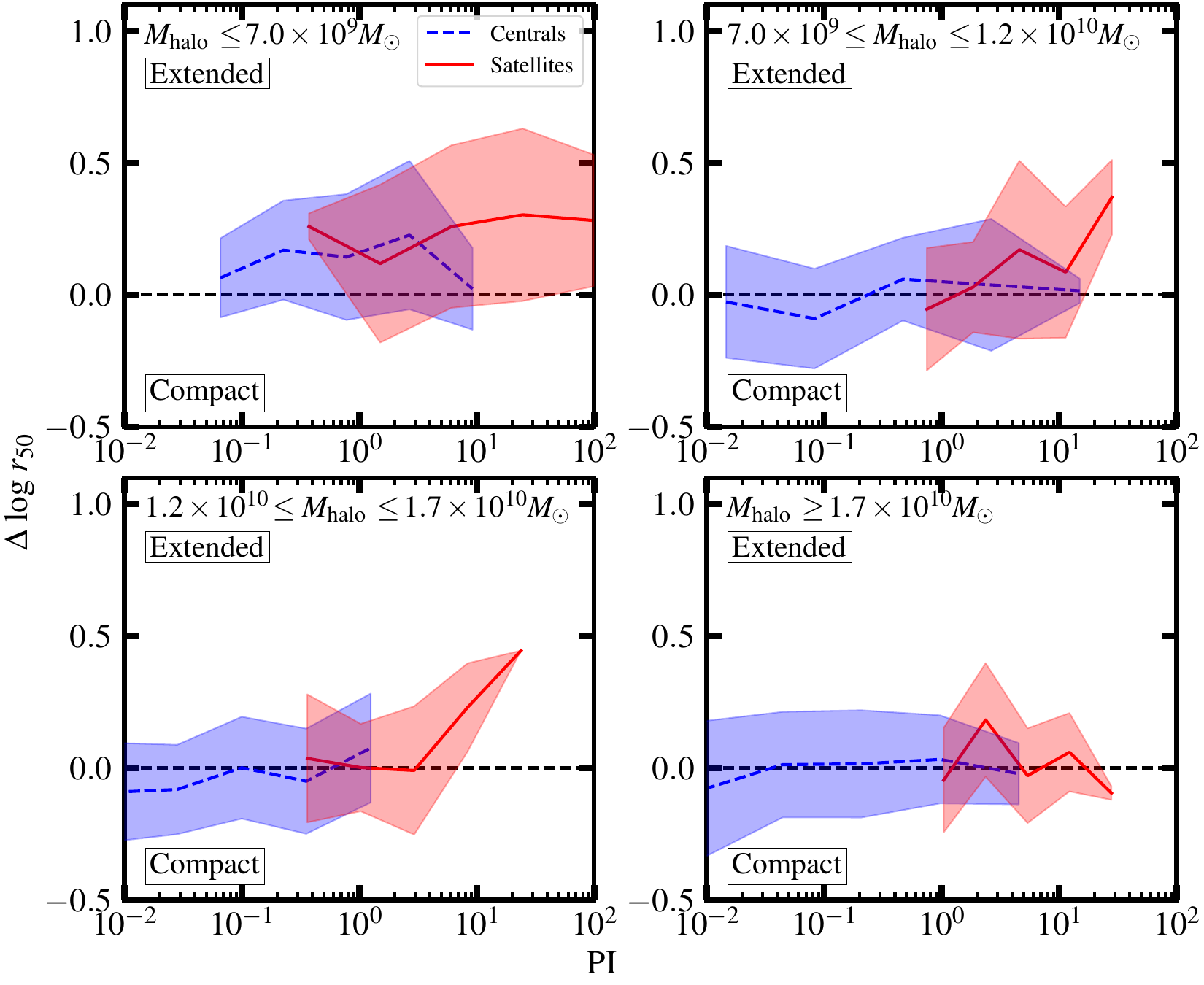}
	\centering
	    \caption{\textit{Environmental impact on relative extent at fixed \mhalo.} Relative extent (\drhalf) as a function of galactic environment (PI) for galaxies divided into four separate halo mass bins, with halo mass increasing from the \textbf{top-left} to the \textbf{bottom-right panel}. As in previous figures, the dashed blue line represents the median behavior of central galaxies, and the solid red line represents that of satellite galaxies. The shaded regions illustrate the 1$\sigma$ scatter about these medians. The horizontal, black dashed like represents the median behavior of the centrals on the SMR. \textit{Takeaway:} In the mass bins where centrals and satellites have substantial overlap, there is a weak positive relationship between \drhalf\ and PI.} 
     \label{fig:Deltar_v_PI_mvirbins}
\end{figure*}

Figure \ref{fig:Deltar_v_PI_mvirbins} shows the relative extent (\drhalf) as a function of galactic environment (PI) for galaxies sorted into four halo mass bins, arranged with increasing mass from the top-left to the bottom-right panel. These bins were chosen to ensure a substantial number of both central and satellite galaxies in each. The dashed blue lines trace the median behavior of centrals, while the solid red lines do the same for satellites, with shaded regions indicating the 1$\sigma$ scatter about these medians. The horizontal black dashed line marks the median centrals’ relation on the SMR. By examining \drhalf\ – PI trends within fixed halo mass bins, we aim to separate the roles of environment and mass, determining whether environment still influences a galaxy’s relative extent once we have accounted for mass.

In the two intermediate mass bins (top-right and bottom-left panels), for galaxies with PI values greater than unity, we find a weak positive correlation between \drhalf\ and PI. These mass ranges are also where centrals and satellites have the greatest overlap, potentially allowing mild environmental effects to emerge more clearly. At the lowest masses (top-left panel), all galaxies appear more extended regardless of environment, likely due to their shallow gravitational potentials, making them inherently more susceptible to internal and external processes (e.g., tidal stripping) that enlarge their apparent sizes. Conversely, at the highest masses (bottom-right panel), all galaxies roughly follow the centrals’ median behavior, suggesting that deep gravitational wells may dominate over environmental influences, diminishing the effect of environment in these massive systems.

\subsection{The complex interplay between environment and mass}\label{sec:interplay}

The results in Figures 4–8 indicate that a galaxy’s relative extent (\drhalf) is governed by a complex interplay between environmental factors and halo mass, while its relative inner dark matter content (\ddm) is primarily determined by halo mass. Additionally, at fixed stellar mass, the environment significantly influences a galaxy’s halo mass. This implies that environmental factors shape galaxy sizes and dark matter distribution not only directly, but also indirectly through their impact on halo mass. In short, these findings highlight a fundamental relationship between environment and halo mass, underscoring the environment’s pivotal role in galaxy formation and evolution.

\begin{figure*}
	\includegraphics[width=1.0\textwidth, trim = 0 0 0 0]{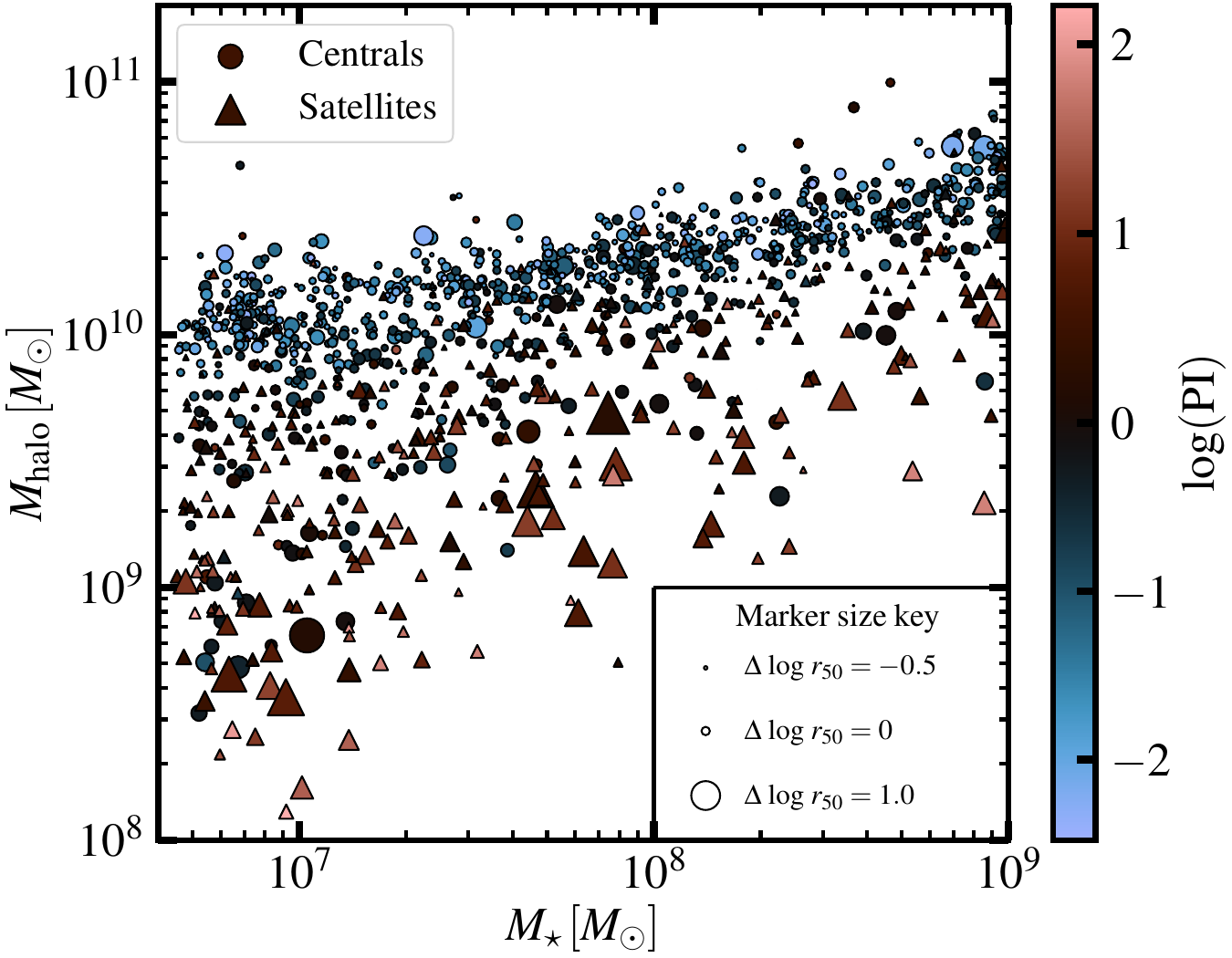}
	\centering
	    \caption{\textit{Relative effects of environment and mass.} Halo mass, \mhalo, as a function of stellar mass, \mstar, color-coded by the logarithm of each galaxy's PI. We scale marker sizes to \drhalf\ to illustrate the relative size of each galaxy compared to others with similar stellar mass. \textit{Takeaway:} Galaxies that live in undermassive halos and in crowded environments are often larger relative to other galaxies of similar stellar mass. However, these are not the only conditions that can yield a relatively extended galaxy.}
     	\label{fig:PI_v_mvir}
\end{figure*}

We illustrate this interplay in Figure \ref{fig:PI_v_mvir}, which shows the \mhalo\ – \mstar\ relation, and scale each marker by \drhalf. Larger markers represent galaxies with greater relative extents (lying well above the median size–mass relation), while smaller markers indicate more compact galaxies. The points are color-coded by the logarithm of the PI value, with red hues highlighting highly perturbed galaxies and blue hues denoting less disturbed ones. This figure demonstrates that a galaxy’s relative size (at fixed stellar mass) depends on both halo mass and environment. For instance, a galaxy situated in an undermassive halo and subject to significant environmental perturbations will appear more extended at fixed stellar mass, as demonstrated by the large, red markers at the lower end of the \mhalo\ – \mstar\ relation. Conversely, a galaxy of similar stellar mass may be more compact if it resides in a more massive halo and experiences minimal perturbation, as represented by the small, blue markers at the upper end of the relation. These contrasting scenarios highlight how both mass and environment jointly influence galaxy structure. However, these two scenarios do not always yield extended or compact galaxies (large, blue circles and small, red triangles.)


\section{Summary and Conclusions}\label{sec:conclusion}

In this paper, we explore how both intrinsic galactic properties and external environmental factors shape galaxy size and dark matter content at fixed stellar mass, using roughly 1,200 low-mass galaxies from the FIREbox cosmological volume simulation \citep{Feldmann2022}. By employing a Random Forest regression analysis, we determine the importance of three features -- halo mass (\mhalo), the stellar-to-halo mass ratio (\mstar/\mhalo), and Perturbation Index (PI) -- in determining how far above or below a galaxy lies on three scaling relations of interest: the \rhalf\ - \mstar\ relation, the \mdmhalf\ - \mstar\ relation, and the \mhalo\ - \mstar\ relation.

Through our Random Forest models, we find that a galaxy’s relative extent (\drhalf) depends on a complex interplay between halo mass (\mhalo), the stellar-to-halo mass ratio (\mstar/\mhalo), and environmental perturbations (PI). In contrast, the relative inner dark matter content (\ddm) is predominantly influenced by \mhalo, underscoring the fundamental role of halo mass in setting the inner dark matter distribution within galaxies.

By examining these relationships across multiple data splits (500 iterations of training and test) we find that the feature importances were stable and robust. This consistency reaffirms that environment, while not the sole driver, is a vital factor shaping galaxy sizes and inner dark matter content, particularly at lower halo masses and for galaxies more strongly perturbed by their surroundings. Additionally, our results show that other features play comparatively smaller roles, and that, at fixed stellar mass, the influence of environment can be overshadowed by mass-based factors as halo masses become large. 

Past work points to halo mass as the fundamental driver of galactic structure, with environment often considered to play a secondary role \citep[e.g.,][]{Kauffmann2004,Zheng2007}. Our results present a refined picture: while halo mass remains a crucial factor in predicting galaxy size and inner dark matter content, the environment, as measured by the PI, has a direct influence on these properties as well as an indirect impact through it's effect on halo mass. These results hint at a complex interplay between the effects that halo mass and environment have on galactic structure. In particular, low-mass galaxies -- whose shallow gravitational potential wells make them more susceptible to external forces -- may undergo more notable changes due to tidal stripping, mergers, and other environmental processes that alter their mass distribution and structural properties \citep{BK2006,Kaufmann2007,Donghia2010,Trujillo2011,Yoon2017}.

Our results suggest that while halo mass provides the underlying ``framework'' that holds a galaxy together, the environment acts as the shifting ground beneath that framework -- pushing, pulling and reshaping its edges. Such environmental influences can strip away dark matter from a galaxy's outer regions, leaving it more dark matter poor and potentially more extended at fixed stellar mass. 

These findings have significant implications for interpreting observations from future large-scale surveys like the Vera C. Rubin Observatory's Legacy Survey of Space and Time (LSST) and the Nancy Grace Roman Space Telescope. With their unprecedented depth and breadth, these missions will provide us with large statistical samples of low-mass galaxies across a range of environments. Our findings suggest that analyses of these datasets must account for both halo mass and environment in order to fully understand the scatter about these scaling relations. Moreover, these rich datasets, when combined with complementary spectroscopic and lensing data, will enable a more nuanced examination of the interplay between halo mass, environment, and galaxy evolution. In doing so, these and other similar surveys will help refine our theoretical models, test the ubiquity of these relationships, and ultimately yield a more complete picture of how galaxies assemble and evolve within their cosmological context.

Our results highlight how environmental influences, operating indirectly through their effect on halo mass, can alter the inner dark matter distribution. This offers a new perspective on the observed diversity in the inner rotation curves of low-mass galaxies \citep{Oman2015}. Instead of attributing the observed variations solely to intrinsic halo formation histories or baryonic feedback \citep[see][]{Ogiya2011,Pontzen2012,DiCintio2014,Onorbe2015,Chan2015,Lazar2020,Mercado2024}, we show that environmental differences can yield a wide array of inner dark matter masses. This suggests that a diversity in inner rotation curve shapes in low-mass galaxies may emerge naturally from the interplay between mass and environment.

Finally, our results reinforce the notion that galaxy formation and evolution cannot be fully understood by considering mass or environment alone. Instead, the fundamental processes shaping galaxies are governed by a complex interplay between halo mass and the environmental factors. This perspective provides a framework that upcoming, more expansive surveys can refine, ultimately leading us toward a more comprehensive understanding of galaxy evolution in our complex cosmic ecosystem.


\section*{Acknowledgments}
FJM is funded by the National Science Foundation (NSF) Math and Physical Sciences (MPS) Award AST-2316748. JSB is supported by NSF grant AST-2408246 and NASA grant 80NSSC22K0827. CK is supported by NSF Graduate Research Fellowship Program (GRFP) grant DGE-1839285 and NASA grant 80NSSC22K0827. LN is supported by the Sloan Fellowship, the NSF CAREER award 2337864, NSF award 2307788, and NSF award PHY-2019786 (The NSF AI Institute for Artificial Intelligence and Fundamental Interactions, \hyperlink{http://iaifi.org/}{http://iaifi.org/}). PFH is supported by a Simons investigator award.

We thank Edwin J. Menendez for advising us in selecting colorblind-friendly colormaps to use for our figures. Finally, we sincerely thank the referee for their thoughtful and constructive feedback, which has helped improve this manuscript.

\vspace{5mm}

\software{The functionalities provided by the following python packages played a critical role in the analysis and visualizations presented in this paper: \textbf{\textsc{matplotlib}} \citep{Hunter2007}, \textbf{\textsc{seaborn}} \citep{Waskom2021}, \textbf{\textsc{Py-SPHViewer}} \citep{BenitezLlambay2015}, \textbf{\textsc{NumPy}} \citep{vanderWalt2011}, \textbf{\textsc{scikit-learn}} \citep{Pedregosa2011}, \textbf{\textsc{SciPy}} \citep{Virtanen2020} and \textbf{\textsc{iPython}} \citep{Perez2007}.}



\bibliography{refs}{}
\bibliographystyle{aasjournal}

\end{document}